\documentclass[prl,preprintnumbers,nofootinbib,aps,twocolumn]{revtex4}

\usepackage{epsfig,amsmath,amssymb}
\usepackage{mathrsfs}
\usepackage[usenames,dvipsnames]{color}
\usepackage[pagebackref=false, colorlinks=false]{hyperref}
\definecolor{redish}{rgb}{0.7,0.2,0.0}  % color defined in (r=red,g=green,b=blue) model
\definecolor{bluish}{rgb}{0.2,0.5,0.8}
\hypersetup{linkcolor=redish,          % color of internal links
                  citecolor=blue,        % color of links to bibliography
                  filecolor=magenta,      % color of file links
                  urlcolor=bluish}          % color of the url
\DeclareFontFamily{U}{rsfs}{}         % Formal Script            %
\DeclareFontShape{U}{rsfs}{m}{n}{<5> rsfs5 <6><7> rsfs7          %
  <8><9><10><10.95><12><14.4><17.28><20.74><24.88> rsfs10}{}     %
\DeclareMathAlphabet{\mathfs}{U}{rsfs}{m}{n}                     %

\newcommand{\ba}{\nopagebreak[3]\begin{eqnarray}}
\newcommand{\ea}{\end{eqnarray}}
\newcommand{\bii}{\begin{itemize}}
\newcommand{\eii}{\end{itemize}}

\begin{document}

\title{Bacteria Around an Acoustic Black Hole: Trapping and Frame-Dragging}
\author{Amitava Banerjee}
\email{amitava8196@gmail.com} 
\author{Ratna Koley}
\email{ratna.physics@presiuniv.ac.in} 
\affiliation{Department of Physics, Presidency University, Kolkata 700073, India.}  
\author{Parthasarathi Majumdar}
\email{parthasarathi.majumdar@rkmvu.ac.in}
\affiliation{Ramakrishna Mission Vivekananda University, Belur Math 711202, India}

\begin{abstract}

Motivated by the need to conceive freely-precessing gyroscopes for detecting acoustic frame-dragging predicted recently in rotating acoustic analogue black holes, we report an incipient investigation on the hydrodynamics of {\it nematic active} fluids. With a specific assumption on barotropicity of a nematic fluid, we discern acoustic analogue black hole spacetimes experienced by linear perturbations of the velocity potential. For vanishingly small diffusivity of the active particles, linear perturbations of the active particle concentration reveal a profile with an enhancement close to the acoustic horizon, hinting towards the possibility of partial trapping of active matter by the acoustic black hole. We further show that, as anticipated, the dynamical nature of the orientation (`polarization') of individual particles indeed opens up the possibility of their use as freely-precessing gyroscopes. In addition, inclusion of diffusivity of active particles in the inviscid solvent is shown to lead to a small effective viscosity. Depending on the sign of the diffusion coefficient, this can either yield superfluid-like behaviour, or enhance the net viscosity, of the nematic system. In either situation, acoustic superradiance, theoretically analyzed and experimentally observed recently for mildly viscous standard fluids, is thus predicted to occur for nematic fluids.

\end{abstract}

\maketitle

\noindent{\it Introduction:} Following up on Unruh's brilliant discovery \cite{r33} of acoustic black hole analogues in barotropic, inviscid fluids, and the prediction of {\it observable} Hawking radiation of phonon excitations from such black holes, theoretical and experimental studies of acoustic analogues of kinematic gravitational phenomena, in condensed matter and cold atom systems, has been a robust ongoing activity \cite{r32,r32a}. The recently reported experimental observation of acoustic Hawking radiation \cite{r34} confirms Unruh's original prediction. Further, acoustic superradiance (`superresonance'), predicted initially for inviscid fluids \cite{r49}, \cite{r50} and more recently for mildly viscous fluids \cite{r44}, is already reported to have been observed in water \cite{r37}. Thus, kinematic gravitational phenomena not observed in physical spacetime are now accessible to laboratory experiments as fluid mechanical analogues \cite{r32}, \cite{r32a} !   

Inertial frame-dragging and the Lense-Thirring precession of gyroscopes, well studied in general relativity for rotating black holes \cite{r38}, have been hitherto unobservable in strong gravity situations. The acoustic analogue of this phenomenon has been analyzed very recently for a rotating acoustic black hole \cite{r39} and quantitative predictions given for the acoustic Lense-Thirring precession frequency. However, despite a proposal \cite{r39} for observation of this phenomenon involving an induced quantum spin for the phonons in certain paramagnetic crystals, the conceptual construction of a freely-precessing gyroscope to detect frame-dragging in the fluid black hole has remained elusive. Our Letter addresses this issue through a first-ever exploration of the hydrodynamics of {\it nematic active} fluids {\it within the analogue gravity paradigm}. We generalize Unruh's approach based on linear perturbations, to discern acoustic black hole analogues, and to study the possibility of conceptualizing gyroscopes out of active particles, in such fluids.  

Active motile agents, {\it {e.g.}} planktons, bacteria, artificial microswimmers to fishes and birds, propel themselves in fluid media and thereby interact among themselves and the media itself. Often such active particles are inherently diffusive and carry an orientation (`polarization') due to their asymmetric structure. In such systems, as the manifestation of fluid-particle and particle-particle interactions, there appear a spectacular variety of fascinating collective dynamical phenomena over a broad range of length and time scales \cite{r1,r2,r2b}.  A key hydrodynamic effect in such phenomena, which arises due to the interaction of the swimmers' orientations with the fluid flow, is the alignment of active swimmers with a shear flow. Such a flow may result in shear-viscosity reduction by the forces generated by swimming, leading to possible `superfluid-like' behavior \cite{r3,r4,r5,r6,r7,r8,r9,r9a}.    

Following Unruh, linear perturbations of the nematic fluid equations and the respective continuity equations for the solvent density, the active particle concentration and orientation, around appropriate stationary backgrounds, are carried out systematically. Acoustic black hole analogues are indeed shown to emerge as possible background flows in nematic fluids, provided an assumption is made regarding barotropicity of the nematic system. Remarkably, even for inviscid solvent fluids, active matter diffusivity is seen to directly produce a weak effective viscosity in the equation describing the perturbed velocity potential in such black hole backgrounds, given in terms of the diffusion coefficient. Adjusting the latter thus paves the way for reducing shear viscosity and confirming the possibility of simulating  an active `superfluid', as also the alternative likelihood of enhancing the net viscosity. Whereas analogue gravity studies traditionally rely on inviscid fluids \cite{r32,r32a}, in view of recent work \cite{r44}, \cite{r34} on rotating acoustic black holes in mildly viscous (`inactive') fluids, both situations offer the opportunity for the incipient observation of acoustic superradiance \cite{r49} in nematic fluids.

For vanishingly small diffusivity, the perturbed active matter concentration around a vortex flow corresponding to a rotating acoustic black hole, exhibits an intriguing profile : a sharp enhancement of concentration is manifest, close to the acoustic horizon ! We shall present a heuristic derivation of this phenomenon in the sequel. Such a behaviour is strongly indicative of the possibility of {\it externally} trapping bacteria and other nematic active elements in a fluid by the acoustic black hole, and may have applications in spatial control of bacterial concentrations in biologically active fluids in general. 

Returning to our initial motivation for investigating active fluids, we consider with some approximations, the continuity equation corresponding to linear perturbations in the {\it orientation} of individual active particles. Numerical solution of the perturbation equation yields a {\it dynamical} orientation angle with respect to the direction of the flow, depending on the flow parameters of the rotating acoustic black hole. Such a strongly dynamical (time-dependent) orientation of an individual bacterium, depending on the background flow parameters, makes this active particle a rather suitable candidate for a {\it freely-precessing gyroscope}. Even though technical details are still awaited, we envisage immediate application of this phenomenon (of dynamical orientation of individual active particle) in acoustic black hole spacetimes), to the detection of acoustic frame-dragging and measurement of the acoustic Lense-Thirring frequency. Since no `weak-field' approximation is involved in our analysis, nematic fluids stand to enable reliable laboratory observations of acoustic analogues of strong-gravity frame-dragging phenomena for the first time ! 

\vglue .1cm

\noindent{\it Model of the Active Fluid And Realization of Analogue Gravity:} We begin with the hydrodynamic equations appropriate to a 
{\it nematic} active fluid \cite{r7}. The total system is described by the following variables: fluid velocity ${\bf v}$, density of the fluid solvent $\rho$, density of the active particles $c$ and their local orientation 
${\bf P}$. 
At this point, we do not assume any relative velocity of the active particles with respect to the background fluid.
The background fluid follows the conventional Navier-Stokes equation for inviscid fluids, with the total stress tensor $\sigma_{ij}$ (whose full expression is given in \cite{r7}) now containing contributions generated by the swimming of the active elements as well as the pressure term $-\Pi\delta_{ij}$ 
\begin{eqnarray}
{{(\rho+c)\left(\partial_t+{\bf v}.{\bf \nabla}\right){\bf v} }}&=& {\partial_{j}\sigma_{ij}} \nonumber \\
\Pi = \Pi(\rho,c) &=& \Pi(\rho+c)~, \label{nseq}
\end{eqnarray}
where, the last equation is taken to characterize barotropicity in the nematic fluid in this Letter. In this respect, since the corresponding assumption of an equation of state relating pressure and density of an active fluid is highly nontrivial and debated \cite{r41,r42}, this crucial assumption restricts the classes of active systems that the analysis in this Letter can be relevant to. Under such an assumption, there is a unique speed of sound given by $v_s^2 \equiv d\Pi/d(\rho +c)$. The continuity equation for the solvent fluid is unchanged and is given by
\begin{equation}
 \partial_t \rho+\mathbf\nabla\cdot \left( \rho{\bf v}\right) =0 \label{cont}
\end{equation}
For the active part, we assume the swimmers in a polarized state with high orientational order among the polarization vectors of individual particles. In that situation, as shown in Ref. \cite{r7}, we can assume $| {\bf P}|$ to be a constant, and work with ${\bf p} \equiv {\bf P}/|{\bf P}|$ instead. Then, the hydrodynamic equations for the variables corresponding to the active elements of such a system can be written as  \cite{r7}
\begin{equation}
\partial_t c+\mathbf\nabla\cdot c{\bf v}=\partial_i\left[D_{ij}\partial_j c+\lambda \gamma' u_{kl}p_{k}p_{l}p_{i}\right] \label{bacc}
\end{equation}
and
\begin{equation}
 [\partial_t+{\bf v}\cdot\mathbf\nabla]p_i+\omega_{ij}p_j\\[5pt]
=\delta_{ij}^T\left[\lambda u_{jk}p_k+\kappa\nabla^{2} p_{j}\right] ~\label{pcont}
\end{equation}
with $\gamma',\kappa,\lambda $ constants, $\delta_{ij}^{T}=\delta_{ij}-p_{i}p_{j}$ being the transverse projection operator$~,~u_{ij}  \equiv (\partial_{i}v_{j}+\partial_{j}v_{i})/2~,~ \omega_{ij} \equiv (\partial_{i}v_{j}-\partial_{j}v_{i})/2$. eqn. (\ref{bacc}) is just the continuity equation for the concentration of active particles, appended with an extra term to account for their diffusion in the background fluid with $D_{ij}$ being the effective diffusion tensor governing orientation-dependent diffusion of active nematics and is given by
\begin{equation}
D_{ij} = D(\delta_{ij}-\xi p_{i}p_{j}) ~\label{difc}
\end{equation}
where $D$ is the diffusion constant and $\xi$ is another constant related to how much diffusion is influenced by the alignment of active particles. The parameter $|\lambda|$ controls the tendency of the swimming particles to align to a shear flow in the background fluid. The polarization equation (\ref{pcont}) contains the usual convective derivative for a flowing fluid and terms involving the symmetric and antisymmetric derivatives of the flow velocity ($\omega_{ij}p_j$ and $\delta_{ij}^T\lambda u_{jk}p_k$) accounting for alignment of the active particles to sheared and rotating fluid flows. Finally, the contribution of elastic energy cost of deforming the polarization field of the particles, which arise due to the aligning interactions between individual particles \cite{r7}, is modeled in the last term in right hand side of Eq. (\ref{pcont}). This model of active matter hydrodynamics can be derived phenomenologically or from microscopic models of the active particle dynamics \cite{r7} and as explained above, is quite intuitive.

We proceed to analogue gravity through linear perturbations for this fluid, following ref. \cite{r32}, assuming an irrotational flow with the velocity 
potential being $\psi$, so that $v_{i}=-\partial_{i}\psi $ and $u_{ij}=-\partial_{i}\partial_{j}\psi $ and $\omega_{ij}=0$. We decompose all 
dynamical fields into their time-independent steady backgrounds (denoted by terms with subscript $0$) and {\it linear} fluctuations 
(denoted by terms with subscript $1$)  : ${\bf{v}} = {\bf{v}}_0 + \epsilon{\bf{v}}_1, ~\rho=\rho_{0}+\epsilon \rho_{1},~c=c_{0}+\epsilon c_{1},~
\Pi =\Pi _{0}+\epsilon \Pi _{1},~\psi =\psi _{0}+\epsilon \psi _{1}$, with $\epsilon << 1$. In this analysis, we keep the orientation field at its static background, and neglect the effect of activity in Eq. (\ref{nseq}) at order $\epsilon $.  With these assumptions and approximations, the equation governing the velocity potential $\psi_{1}$ may be written as
  
   \begin{widetext}
  \begin{eqnarray}
  -\partial_{t}\left[\left(\frac{c_{0}+\rho_{0}}{v_{s}^{2}}\right)(\partial_{t}\psi _{1}+{\bf v}_{0}. \nabla \psi _{1})\right]
  &+& \nabla.\left[(c_{0}+\rho_{0})\nabla\psi _{1}-\left(\frac{c_{0}+\rho_{0}}{v_{s}^{2}}\right)(\partial_{t}\psi _{1}
 +{\bf v}_{0}.\nabla \psi _{1}){\bf v}_{0}\right] \nonumber \\ &=& \partial_{i}\left[-D_{ij}\partial_{j}\left(\frac{c_{0}}{v_{s}^{2}}
 (\partial_{t}\psi _{1}+{\bf v}_{0}. \nabla \psi _{1})\right)+
 \lambda \gamma 'p_{k}p_{l}p_{i}\partial_{k}\partial_{l}\psi _{1} \right]~. \label{main}  
 \end{eqnarray}
  \end{widetext}
  One can interpret Eq. (\ref{main}) in terms of the standard analogue gravity acoustic spacetime \cite{r32}. thus the LHS of 
Eq. (\ref{main}) reduces to the covariant d'Alembertian wave operator $\Box_g^2$ corresponding to the acoustic metric given by

\[
g^{\mu \nu }=\frac{1}{(\rho _{0}+c_0)v_{s}}
\left[
\begin{array}{c|c}
-1 & -v_{0}^{j} \\
\hline
-v_{0}^{i} & v_{s}^2\delta ^{ij}-v_{0}^{i}v_{0}^{j}
\end{array}
\right]
\]

 acting on the perturbed potential $\psi_1$. 
The RHS is clearly not Lorentz invariant and thus offers new opportunities to simulate Lorentz-violation Physics \cite{r43}. In particular, neglecting polarization terms in Eq. (\ref{main}) one obtains,

\begin{equation}
\frac{\rho_0+c_0}{v_s^2} \Box_g^2 \psi_1 =\frac{D c_0}{v_s^2} \nabla^2 (\partial_t + {\bf v}_0 \cdot \nabla) \psi_1  \label{waveq}
\end{equation}

where we have used the relation 
\begin{equation} c_1 = (c_0/v_s^2) (\partial_t + {\bf v}_0 \cdot \nabla) \psi_1 \label{c1eqn} \end{equation}
                                                                                   which follows from barotropicity. An interesting aspect of Eq. (\ref{waveq}) is its striking similarity with the wave equation for $\psi_1$ derived in conventional barotropic fluids with a small viscosity $\eta$ \cite{r44}
\begin{equation}
\frac{\rho_0+c_0}{v_s^2} \Box_g^2 \psi_1 = - \frac43 \frac{\eta}{\rho_0 v_s} (\partial_t + {\bf v}_0 \cdot \nabla) \nabla ^2\psi_1  \label{visc}
\end{equation} 
which, modulo some extra terms, implies that nematic fluids have an effective viscosity arising from diffusion, even when the solvent fluid is inviscid. In that respect, it is similar to the known fact \cite{r3,r4,r5,r6,r7,r8,r9,r9a} that flow-aligning swimmers can enhance or reduce the viscosity of the solvent by aligning to a shear flow and generating flow patterns about themselves, reinforcing or opposing the existing shear. On one hand, these effects help impart `superfluidic' property to active systems \cite{r4}, which is usually an essentiality \cite{r32} to realize conventional analogue gravity. On the other hand, these terms contribute as explicitly Lorent-violating terms \cite{r44} in the model, thereby facilitating simulations of the physics of Lorentz violation \cite{r43}. From studies on such systems, one concludes, as demonstrated in \cite{r44}, acoustic superradiance (`superresonance'), i.e., amplification of sound waves at the cost of the rotational energy of the acoustic analogue black hole, is certainly possible for nematic fluids. The physical effects of such phenomena on active systems is left as a possible future work.

\vspace{0.5 cm}

\begin{figure}[h]
\includegraphics[width=5cm, angle=-90]{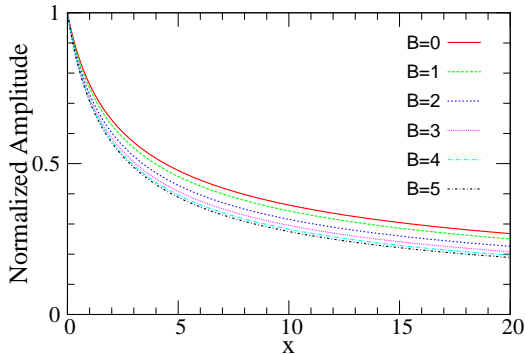}
\hfill
\caption{\label{Fig 1.} Active particle concentration profiles $\frac{|c_{1}(x)|^2}{|c_{1}(0)|^2}$ for $v_s=1,\omega =0.001,m=1,A=1$.}
\end{figure}

\noindent{\it Active Matter Concentrations in Draining Sink Flow:} Since the equation for the perturbed velocity potential in the active fluid is complicated in general, we shall first make several simplifying assumptions. We neglect the contributions from the diffusion of active matter by assuming $D<<L^{2}/\tau$, where $L$ and $\tau$ are the characteristic length and times scales of the system. We also neglect the term $\lambda \gamma 'p_{k}p_{l}p_{i}\partial_{k}\partial_{l}\psi _{1}$ since it contains products of direction cosines. To see the effects of the acoustic geometry, we shall calculate the distribution of the concentration fluctuation of the active particles. As an example of the background flow, we choose the `draining sink' (DS) vortex profile
\begin{equation} 
{\bf v}_{0}=\frac{-A \hat r+B \hat \phi}{r} \label{dseq}
\end{equation}
with $A>0$ and B being two parameters. We also assume a homogeneous $c_0$ and $\rho _0$ profile leading to homogeneous sound speed. This velocity profile is widely used in the analogue gravity community \cite{r32,r32a,r37} and indeed vortical flows are ubiquitous in active matter systems \cite{r15,r21,r45,r46,r47,r48} as well. The corresponding metric is well-known to admit an acoustic trapping horizon at $r=r_{H}=\frac{A}{v_{s}}$ where the inward radial component of the fluid velocity equals sound speed and to demonstrate superresonance \cite{r49,r50} and inertial frame dragging \cite{r39}. To calculate the fluctuation in the active matter concentration $c_{1}$, we shall use an explicit solution of the homogeneous wave equation, which is valid near the horizon and given by \cite{r49,r50}
\begin{equation}
 \psi_{1m}(r,\phi,t)=e^{i(m\phi-\omega t)}e^{\frac{imB}{2A}\ln(1+x)}F(x)
 \end{equation} 
where $m\in I^{+}$ is the mode number, $\omega >0$ is the frequency of the mode, $x=\frac{r^2}{r_{H}^2}-1$ is the normalized squared 
radial distance from the horizon and $F(x)={}_2F_{1}(\alpha,\beta;\gamma;-x)$ is the hypergeometric function with 
$Q=\frac{\omega A}{v_{s}^{2}}-\frac{Bm}{A}, 
S=\sqrt{m^2+\frac{2Bm\omega}{v_s^2}-\frac{2\omega^2A^2}{v_s^4}}$, $\alpha=-\frac{S}{2}-i\frac{Q}{2}, \beta=\frac{S}{2}-i\frac{Q}{2}, \gamma=1-iQ$. This solution is valid outside the horizon and in the low-frequency 
regime where the relations $\frac{\omega A}{v_{s}^2}x<<m$ and $\frac{\omega A}{v_{s}^2}<<1$ hold. Using Eq. (\ref{c1eqn}), we see that $c_{1}$ has a wave-like fluctuation in time and angular coordinate, with an amplitude 
which is a function of the radial distance from the centre given by the equation, 
\begin{equation}
|c_{1}|^2\sim  \left|i\omega F(x)+2\frac{v_s^2}{A}\frac{dF}{dx}\right|^2
\end{equation}
which is plotted in Fig. \ref{Fig 1.}. 
%%%%%%%%%%%%%%%%%%%%%%%%%%%%%%%%%%%%%%%%%%%%%%%%%%%%%%%%%%%%%%%%%%%%%%%%%
%%%%%%%%%%%%%%%%%%%%%%%%%%%%%%%%%%%%%%%%%%%%%%%%%%%%%%%%%%%%%%%%%%%%%%%%%%%%%%
The plot reveals that there is an enhancement of the amplitude $|c_{1}|^2$ near the horizon, and a corresponding monotonic decay away from it. Furthermore, we also see that the length scale of the radial decay of this oscillation amplitude can be controlled by the parameter $B/A$ of the background flow. In general, this spatial variation of the active matter density wave amplitude comes as a direct consequence of the underlying acoustic metric which controls the perturbation in the velocity potential $\psi_{1}$. As seen from Eq.(\ref{c1eqn}), this variation can be controlled totally by manipulating the background fluid flow ${\bf{v}}_0$. As such, this be useful in localizing bacteria or colloidal particles or in targeted delivery of such objects with a carefuly designed flow. This general scheme of controlling amount of fluctuation in active matter concentration will complement the existing methods of trapping and manipulation of active particles using, for example, topological defects in the orientation profiles \cite{r51,r52,r52b}, curvature \cite{r52a}, deformation \cite{r53}  or constriction \cite{r54} of the bounding walls, swirl flows \cite{r15,r55} or shear flows \cite{r56,r57}, most of which are very system-specific. 
\vglue .1cm
\noindent{\it Active Nematic Particles as Probes of Acoustic Lense-Thirring Precession:} Rotating acoustic black hole spacetimes like that realized in DS offer the possibility of observation of acoustic inertial frame dragging and Lense-Thirring precession \cite{r39}, wihtout resorting to any weak field approximation. In this respect, active particles in the vortex flow can work as freely-rotating gyroscopes, probing acoustic frame-dragging and measuring the corresponding Lense-Thirring precession frequency. This would be best observable with extremely dilute suspensions of active particles where we can neglect self-interactions of the orientation field. In such limits, we can set $c_0=0$ and thus the acoustic geometry has no contributions from particle concentration. Then, we can consider the effect of the fluid flow on the orientation ${\bf p}$ and centre-of-mass position ${\bf x}$ of the individual particles and write their equations of motion as: 
\begin{equation}
 \dot {\bf x}={\bf v}
 \end{equation} 
which implies that the center of mass of each particle drifts with the same speed as the fluid flow while their orientations tends to align to local shear in the flow according to the Jeffrey's equation\cite{r58} corresponding to an irrotational flow 
\begin{equation}
 \dot p_{i}=\beta u_{jk}p_k(\delta_{ij}-p_{i}p_{j})
 \end{equation} 
where $\beta$ is a parameter depending on the shape and aspect ratio of individual active particles. If we write ${\bf p}=(\cos\theta, \sin\theta)$ then, using Jeffrey's equation this leads to the equation for orientation angle $\theta$ as
\begin{equation}
\partial_{t}\theta=\beta\left[u_{xy}\cos2\theta-\frac{u_{xx}-u_{yy}}{2}\sin2\theta \right]~. \label{thet} 
\end{equation}
We numerically solve this equation for the DS velocity profile we used above and plot the orientation fluctuation due to the velocity perturbation in the fluid. 
%%%%%%%%%%%%%%%%%%%%%%%%%%%%%%%%%%%%%%%%%%%%%%%%%%%%%%%%%%%%%%%%%%%%%%%%
\begin{figure}[h]
{\bf (a)}\includegraphics[width = 5 cm,angle = -90]{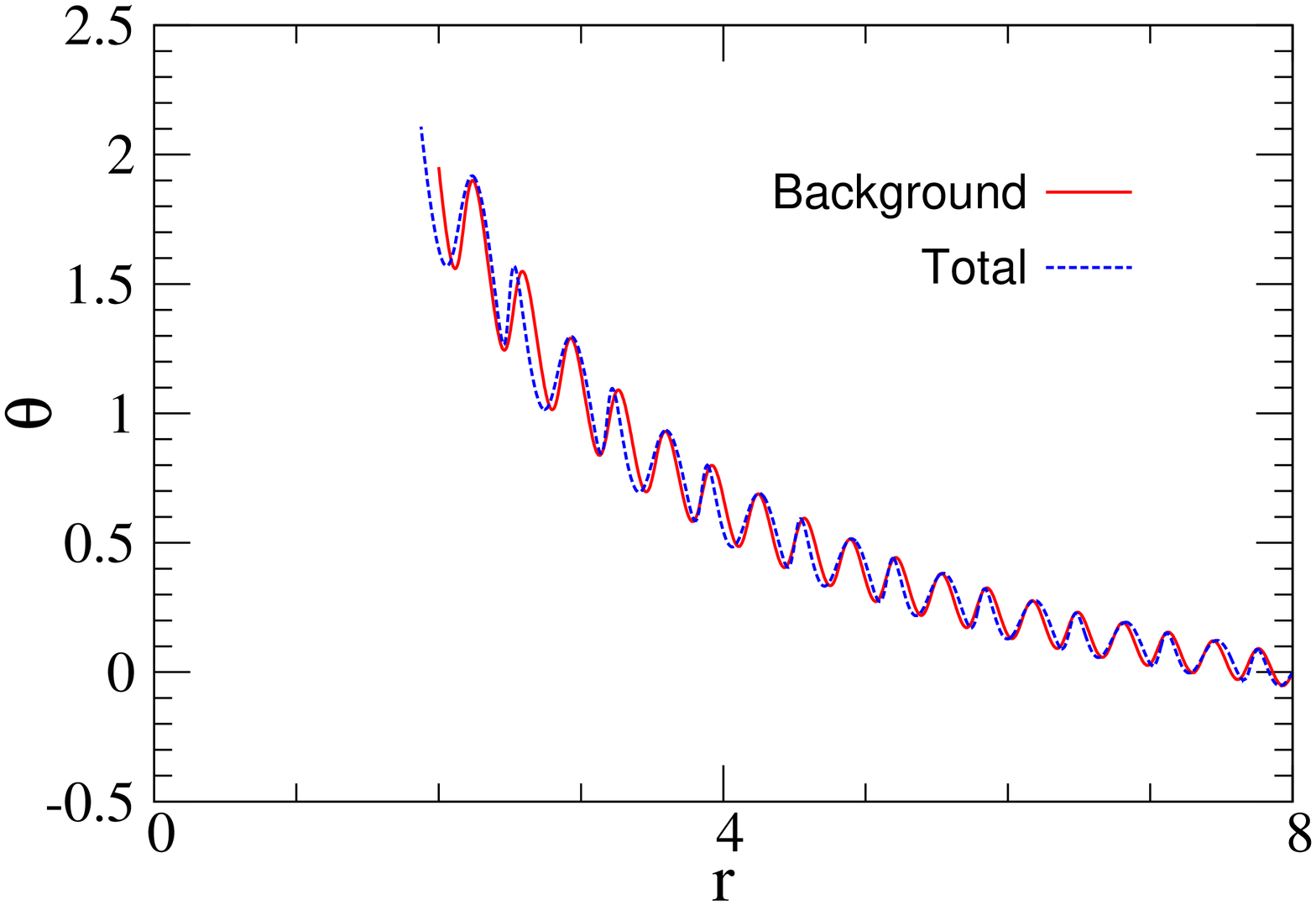}\\
{\bf (b)}\includegraphics[width = 5 cm,angle = -90]{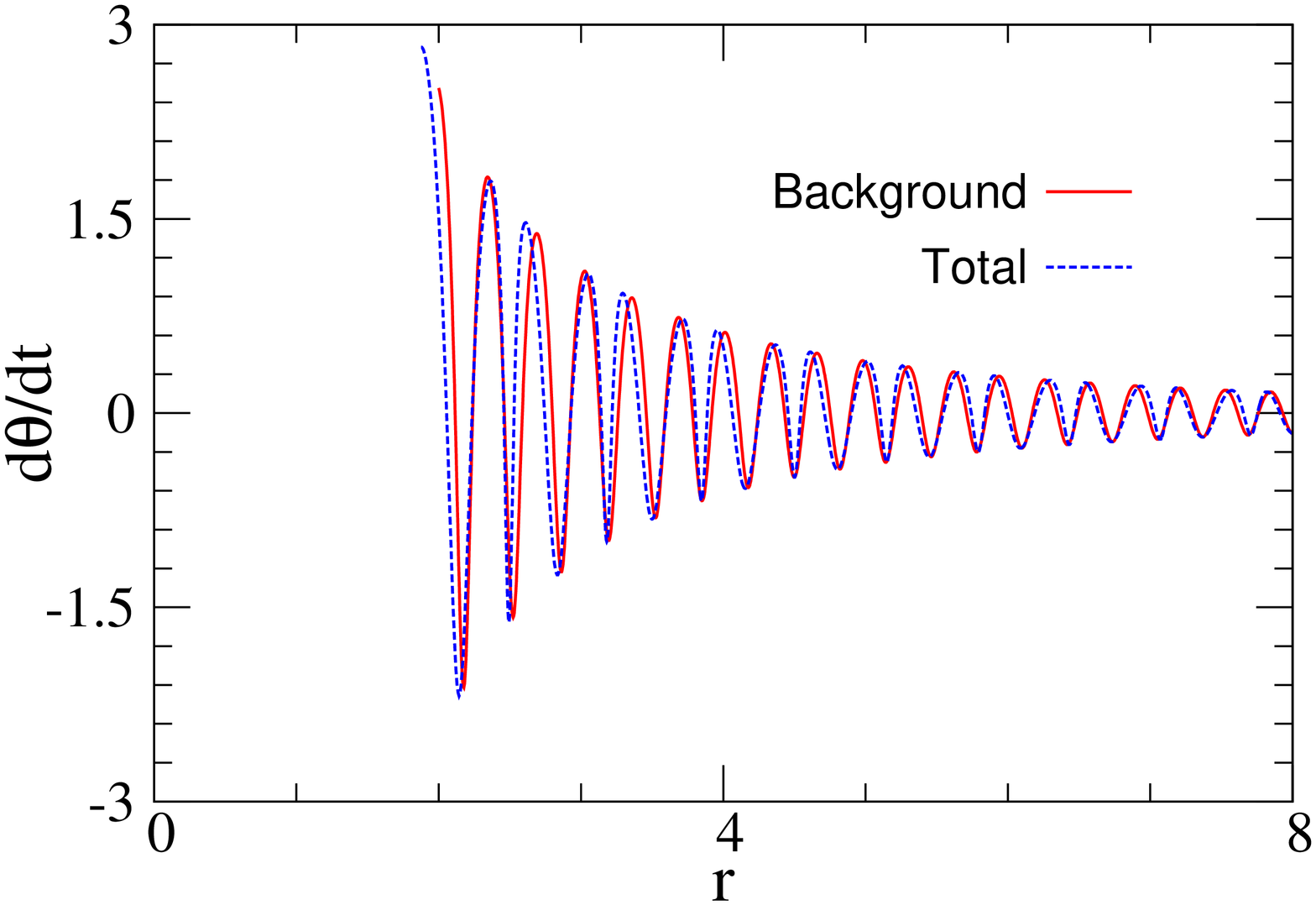}
\caption{\label{Fig 2.} Dynamics of the (a) orientation $\theta $ and (b) angular velocity $d\theta /dt$ of active particles with radial coordinate $r$; with $\omega =0.001,v_s=1,m=1,A=1,B=10,\beta =1,\epsilon =0.1$. The initial conditions are $x=8,y=0,\theta =0$, the `background' corresponds to the dynamics with ${\bf v}={\bf v_0}$ while the case with ${\bf v}={\bf v_0}+\epsilon Re[{\bf v_1}]$ is plotted as `total'.}
\end{figure}
%%%%%%%%%%%%%%%%%%%%%%%%%%%%%%%%%%%%%%%%%%%%%%%%%%%%%%%%%%%%%%%%%%%%%%%%%%%%
The resulting Fig. \ref{Fig 2.} shows that the temporal fluctuations in the orientation due to $\psi _1$, representing precession, are particularly prominent for larger $B/A$ ratio and are enhanced as the particles approach the acoustic horizon. The former observation can be attributed to the fact that in our acoustic metric, the angular momentum of the black hole horizon $\Omega_H=(B v_s^2)/A^2$ \cite{r39} increases with $B/A$ and frame-dragging effects increase accordingly. This precession of the orientation of individual active particles can be experimentally observed directly. The above simulation results can thus be compared with those observations possibly after averaging over the dynamics of a sufficiently large number of individual particles to eliminate random fluctuations in their position and orientations resulting in translational and rotational diffusion. The effects would be best observable for very small rotational diffusion constant ($D_r$) values relative to the orientation fluctuations $\delta\theta$ due to 
${{\bf v}_1}$, i.e., $2 D_r \tau <<\delta\theta^2$ where $\tau$ is the time needed to reach the horizon from where the particle starts. In those cases, we can use the dynamics of active particles to probe the corresponding dynamics of test objects in the emergent acoustic spacetime.
\vglue .1cm
\noindent{\it Concluding Remarks:} In the foregoing, active nematic fluids have been shown to emerge as a novel arena for analogue gravity phenomena. We have demonstrated several implications of this symbiotic relationship between active fluids and acoustic gravity analogues, working within certain approximations for simplicity. With regard to future outlook, inclusion of the contribution of the diffusion term in the right hand side of our wave equation, neglected in our analyses so far, is a high priority. While ignoring this contribution is reasonable for the limits of low concentration or low diffusivity which we mentioned earlier, inclusions of such terms may account for simulations of more exotic phenomena. Furthermore, for small $D$, the effect of the source term in the wave equation can be evaluated perturbatively starting from a solution of $\psi _1$ for the corresponding homogeneous wave equation. Analysis of such kinds is inspired by the fact that since the diffusion (by eqn.(\ref{difc})) is dependent on the polarization of the particles, it can be controlled experimentally if we can manipulate the alignment of the particles by, e.g., the fluid flow itself as we discussed, passive liquid crystals \cite{r59}, intense laser beams \cite{r60} or external magnetic fields for magnetotactic bacteria \cite{r61}. This would allow us greater control over the dispersion of the active particles over the fluid, complementing our proposed way of doing the same by imposing analogue spacetime structures to the flow perturbations. In summary, since both active matter and analogue gravity enjoy established stature as robust areas of reseach in contemporary theoretical and experimental sciences, this incipient investigation of analogue gravity in active nematic fluids offers myriad possibilities of rich and novel phenomena involving both biology and physics.

\vglue .1cm
\noindent{\it Acknowledgements:} We thank O. Ganguly and S. Chatterjee for useful discussions. One of us (PM) thanks C. Chakraborty, L. Crispino and J. Kunz for illuminating remarks and discussions.

\end{document}